\documentclass[11pt]{article}  %Do not change the point size.
\usepackage{menuproc}
\usepackage{cite}
\usepackage{epsfig}
\usepackage{amsmath,amssymb}
\begin{document}
%
%____________________________________________________________
%
%  Title, authors, institutions, and abstract
%----------------------------------------------------------------
%  Syntax:  \titlematter{title}{authors}{institutions}{abstract}
%----------------------------------------------------------------
%     If lines are too long, use linebreaks where convenient.
%     If all authors are from the same institution, omit raised letters.
%
\titlematter{The pion-nucleon $\Sigma$ term is definitely large:\\
  results from a G.W.U. analysis of $\pi N$ scattering data}%
{M.M. Pavan$^a$, R.A. Arndt$^b$, I.I. Strakovsky$^b$ and R.L. Workman$^b$}%
{$^a$University of Regina\\ TRIUMF, Vancouver, B.C. V6T-2A3, Canada\\
 $^b$Center for Nuclear Studies, Department of Physics,\\
 The George Washington University, Washington, DC 20052, U.S.A.}%
{A new result for the $\pi N$ $\Sigma$ term from a George Washington
  University/TRIUMF group analysis of $\pi N$ data is presented. The value
  $\Sigma$=79$\pm$7 MeV was obtained, compared to the canonical value
  64$\pm$8 MeV found by Koch. The difference is explained simply by the PSI
  pionic hydrogen value for a$_{\pi ^{-}p}$, the latest results for the
  $\pi NN$ coupling constant, and a narrower $\Delta$ resonance. Many
  systematic effects have been investigated, including Coulomb corrections,
  and database changes, and our results are found to be robust.  In the
  standard interpretation, our value of $\Sigma$ implies a nucleon
  strangeness fraction y/2$\sim$0.23.  The implausibility of such a large
  strange component suggests that the relationship between $\Sigma$ and
  nucleon strangeness ought to be re-examined.}
%
%
%____________________________________________________________
%  Start article here:

%%%%%%%%%%%%%%%%%%%%%%%%%%%%%%%%%%%%%%%%%%%%%%%%%%%%%%%%%%%%%%%%%%%%%%%%%%%%%%%%%%
\section{Introduction}

The $\pi N$ sigma term $\Sigma$ has long been a thorn in the side of low
energy quantum chromodynamics (QCD)~\cite{reya74,jaff80}. The canonical
result $\Sigma$=64$\pm$8 MeV was obtained by Koch~\cite{koch,bible} based
on an analysis of pre-1980 $\pi p$ and $\pi \pi$ scattering data,
KH80~\cite{bible,kh80}.  Gasser, \textit{et al.}~\cite{glls88} later
developed an alternative method of extracting $\Sigma$ which agreed
perfectly with Koch when using the same KH80 solution. In the usual picture, the
nucleon strangeness parameter is
\begin{equation}
  y/2 = \frac{<N|\overline{s}s|N>}{<N|\overline{u}u+\overline{d}d|N>}
\end{equation}
The canonical $\Sigma$ result yields $y=0.11\pm0.07$, whereby the strange
quarks would contribute $\sim$110 MeV to the nucleon mass, an amount
considered too large to be physical in light of results from {\it e.g.}
neutrino scattering~\cite{jaffe-ref}. This ``sigma term puzzle'' spawned a
whole generation of $\pi N$ scattering experiments that have greatly
increased the size and the quality of the scattering database.

A long-standing prejudice has been that new and better $\pi N$ scattering
data and an updated analysis ultimately would result in a smaller value for
$y$. With the new generation of experiments almost all completed, our
George Washington University/TRIUMF group has sought to extract the \(
\Sigma \) term as part of our ongoing $\pi N$ partial-wave and dispersion
relation analysis program, which employs the most up-to-date $\pi N$
scattering data in our SAID database\cite{said}.  Our main conclusion is
that contrary to wishful expectation, a thorough analysis of the new data
has yielded a {\it larger} value, $\Sigma=79\pm7$ MeV, which can be
understood simply in light of the new experimental information.
The sigma term and our analysis will be summarized briefly. Details
can be found \textit{e.g.} in Refs.~\cite{bible,glls88,vpiPapers,pav99}.

%--------------------------------------------------------------------------------
\section{The Pion-Nucleon Sigma Term}

%--------------------------------------------------------------------------------
\begin{figure}[t]
\parbox{.55\textwidth}{\epsfig{file= 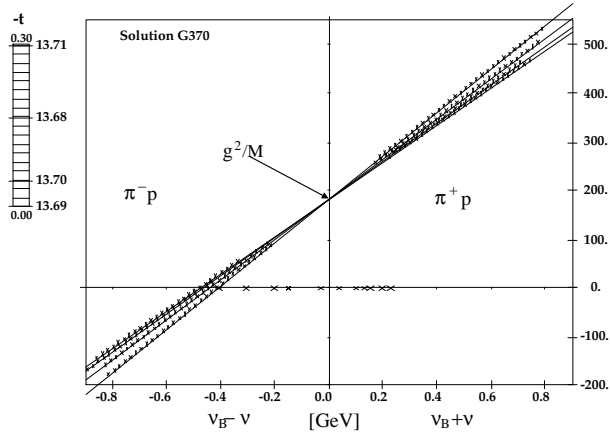,width=.5\textwidth,silent=,clip=}}
\hfill
\parbox{.4\textwidth}{
  \caption{
    \small Determination of the $\pi NN$ coupling constant from the H\"uper
    dispersion relation. The y-intercept gives the coupling \protect\(
    g^{2}/M\protect \), and the left (right)-hand side of the figure is
    dominated by \protect\( \pi ^{-}p\protect \) (\protect\( \pi
    ^{+}p)\protect \) data. This technique is well suited to determine the
    coupling constant since most systematic effects ({\em e.g.} Coulomb
    corrections) affect each side asymmetrically, ``pivoting'' the curves
    about the intercept, hence greatly reducing their effect on $g^2$.
    \normalsize}
  \label{fig:huper}
}
\end{figure}

The sigma term $\hat{\sigma}$ measures the nucleon mass shift away from the
chiral ($m_{u}=m_{d}=0$) limit, thereby parameterizing the explicit
breaking of chiral symmetry in QCD due to the non-zero up and down quark
masses. Models of nucleon structure are required to determine
$\hat{\sigma}$. The canonical result $\hat{\sigma}=35\pm5$ MeV is due to
Gasser\cite{gas81} based on SU(2) chiral perturbation theory plus meson
loop corrections. One obtains the strangeness $y$ from
\begin{equation}
  \sigma(0) = \frac{\hat{\sigma}}{1-y}
  \label{eqn:sigmahat}
\end{equation}
where the theorem of Brown, Pardee, and Peccei\cite{brow71} relates
$\sigma(0)$ to the isoscalar invariant  $\pi N$ scattering
amplitude $D^{+}(\nu,t)$ at the ``Cheng-Dashen''
point\cite{chen71}, $\nu=0, t=2m_{\pi}^2$:
\begin{eqnarray}
  \Sigma & = & F_{\pi}^2 \bar{D}^{+}(0,2m_{\pi}^2) \nonumber \\
         & = & \sigma(2m_{\pi}^2) + \Delta_R
\end{eqnarray}
where
\begin{equation}
  \sigma(2m_{\pi}^2) = \sigma(0) + \Delta_{\sigma}
\end{equation}
and $F_{\pi}$=92.4 MeV is the pion decay constant, $\nu$ is the crossing
energy variable, and $t$ is the four-momentum transfer. The ``remainder
term'' $\Delta_R$ is small ($<$2 MeV\cite{bern96}). The nucleon scalar form
factor $\sigma(t)$ shifts by an amount $\Delta_{\sigma}$=15 MeV from $t=0$
to $ t=2m_{\pi}^2$, calculated from a $\pi\pi$ dispersion relation
analysis\cite{glls88} and recently confirmed by a chiral perturbation
theory calculation\cite{bech99}. The bar over $\bar{D}^{+}$ indicates that
the pseudo-vector Born term has been subtracted.

The Cheng-Dashen point lies outside the physical $\pi N$ scattering region,
so the experimental $\bar{D}^{+}$ amplitude must be extrapolated to obtain
$\Sigma$.  The most reliable extrapolations are based on dispersion
relation (DR) analyses of the scattering amplitudes~\cite{bible}.  The
Koch result $\Sigma = 64\pm8$ MeV was based on hyperbolic dispersion
relation calculations~\cite{koch}. More recently, Gasser, \textit{et al.},
(GLLS)~\cite{glls88} developed another dispersion theoretic approach based
on forward subtracted $\pi N$ dispersion relations. Expanding $D^{+}(t)$ as
a power series in $t$, the experimental sigma term $\Sigma$ can be
expressed as
\begin{eqnarray}
  \Sigma &=& F^2_{\pi}(\bar{d}^+_{00}+2m^2_{\pi}\bar{d}^+_{01}+\ldots)\\
         &=& F^2_{\pi}(\bar{d}^+_{00}+2m^2_{\pi}\bar{d}^+_{01}) + \Delta_D \\
         &=& \Sigma_d +  \Delta_D
\end{eqnarray}
The GLLS, or truncated, sigma term $\Sigma_d$ is obtained via the
subthreshold coefficients $\bar{d}^{+}_{00}$ and $\bar{d}^{+}_{01}$,
calculated from the forward subtracted $\bar{D}^{+}$ and ``derivative''
$\bar{E}^{+}$ dispersion relations, respectively. They can also be
determined from the subtraction constants $D^{+}(0,t)=C^{+}(0,t)$ in the
fixed-t dispersion relation $C^{+}(\nu,t)$. The intercept of the curve
$D^{+}(0,t)$ yields $d^+_{00}$, whereas the slope at $t=0$ yields
$d^+_{01}$. The ``curvature correction'' term $\Delta_{D}=12\pm1$ MeV was
determined from a $\pi\pi$ dispersion relation analysis~\cite{glls88}. The
great advantage of this approach is that $\sigma(0)$ can be obtained simply
from $\Sigma_d$ via\cite{glls88}
\begin{equation}
  \sigma(0) =\Sigma_d - (3\pm3) \mbox{MeV}
\end{equation}
since the correction terms $\Delta_{\sigma}$ and $\Delta_D$ almost cancel,
both having similar $\pi\pi$ amplitude input~\cite{glls88}.

The analysis of Ref.\cite{glls88} used the Karlsruhe KH80~\cite{kh80} $\pi
N$ phases as input and fit just  the low energy data. Their result was
$\Sigma_{d}\sim$50 MeV, or $\Sigma\sim$62 MeV (with $\Delta_{D}$=12 MeV), in
agreement with Koch~\cite{koch}. Questions regarding the accuracy of
the $E^+$ dispersion relation integral, which is more sensitive to the
smaller and more poorly known higher partial waves than other dispersion
relations, were answered by the good agreement which demonstrated the
reliability of the approach.

\section{Analysis Procedure}
Solutions from our ongoing $\pi N$ partial-wave and dispersion relation
analysis are released when changes to the database and analysis method
warrant~\cite{said}. Details of our analysis method can be found in
Ref.~\cite{vpiPapers,pav99,pav99b}. An energy-dependent $\pi N$
partial-wave analysis (PWA) is performed on the available data up to 2.1
GeV pion laboratory kinetic energy, applying constraints from forward
$C^{\pm}(\omega)$ and $E^\pm(\omega)$ DRs, as well as fixed-t
$B_{\pm}(\nu,t)$ (in the ``H\"uper" form~\cite{bible}) and $C^{\pm}(\nu,t)$
DRs.  These dispersion relations are constrained~\footnote{This range has
  been increased from previous analyses, and in practice the dispersion
  relations are well satisfied somewhat beyond that range due to the energy
  dependent partial wave forms} to be satisfied to within $<$2\% from 30 to
800 MeV for $-0.4 < t < 0.0$ GeV${^2}/c{^2}$. The dispersion integrals use
the Karlsruhe KH80 phases from 2.1 to 4.5 GeV and high energy
parameterizations above that using forms found in Refs.~\cite{bible,mar70}.

Dispersion relations depend on {\it a priori} unknown constants
\textit{e.g.} scattering lengths and $g^2$.  Our analysis determines these
constants by a best fit to the data and the dispersion relations. The
coupling $g^2$, the $\pi^-p$ s-wave scattering length $a_{\pi^-p}$, and the
p-wave scattering volume $a^{+}_{1+}$ were fixed for each fit over a grid
of values (for reasons of fit stability), where the combination with the
lowest $\chi^2$ yields the final solution. The fitting procedure
automatically chooses the best-fit isovector scattering length $a^{-}_{0+}$
and volume $a^{-}_{1+}$, and the subtraction constants $C^{\pm}(0,t)$. This
method enables us to check their sensitivity to various systematic effects,
\textit{e.g.} database changes.

The low energy $P_{13}$ partial wave is constrained to follow the expected
partial wave dispersion relation behaviour in its Chew-Low approximation
form\cite{ham64}, which our other p-waves satisfy without constraint. As
well, the low energy F and higher partial waves, too small to be
determined from the $\pi N$ scattering data, are constrained to agree with
those calculated by Koch\cite{koch86} from partial wave projections of
fixed-t dispersion relations, which are dominated by t-channel ($\pi\pi$)
contributions.  This ensures that our higher partial waves satisfy
analyticity and unitarity requirements.

%--------------------------------------------------------------------------------
\section{Results}

\begin{figure}[t]
\parbox{.52\textwidth}{\epsfig{file=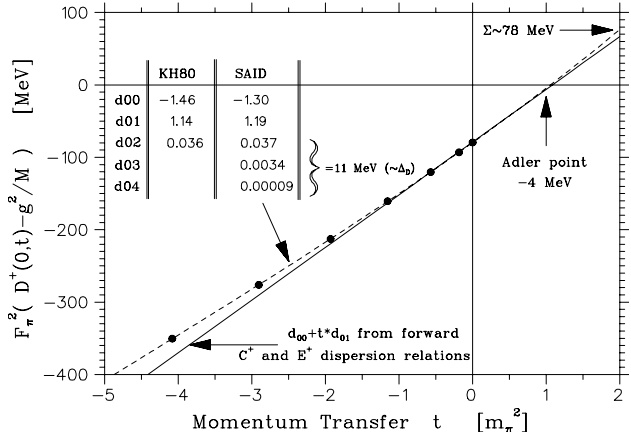,width=.5\textwidth,silent=,clip=}}
\hfill
\parbox{.45\textwidth}{
  \caption{\small
    The amplitude $\overline{C}^{+}(0,t)$ (points) evaluated from fixed-t
    $C^{+}(\nu,t)$ dispersion relations.  A fit (dashed line) yields the
    subthreshold coefficients in the table. The solid diagonal line is
    inferred from forward $C+$ and $E^+$ dispersion relations, and agrees
    perfectly with $\overline{d}_{00}$ and $\overline{d}_{01}$ in the
    table.  The curvature terms ( $\overline{d}_{0i}, i\geq2$) imply
    $\Delta_{D}>$11 MeV, consistent with the canonical result 12$\pm$1 MeV
    from Ref.~\protect\cite{glls88}.  The amplitude is very small as
    expected at $t=m_{\pi}^2$ (``Adler point'').  The overall consistency
    tends to support our result for the sigma term, $\Sigma\sim79\pm7$ MeV.
    \normalsize}
\label{fig:c+Extrap}
}
\end{figure}

Our main results are summarized in Figs.~\ref{fig:huper}
and~\ref{fig:c+Extrap} and Table~\ref{tab:sigma_terms}. We find for the
$\pi NN$ coupling constant $g^{2}/4\pi = 13.69\pm0.07$ (or
$f^{2}=0.0757\pm0.0004$), stable in our solutions for many years
(see~\cite{pav99b}). Our coupling constant agrees with most other recent
results, in particular the comprehensive $NN$ and $\pi N$ analyses of the
Nijmegen group (see Ref.~\cite{deSw97} and references cited therein). Note
that this result is perfectly consistent with both the Goldberger-Treiman
discrepancy~\cite{gol58,dom85} and the Dashen-Weinstein sum
rule~\cite{goit99}, removing a long-standing inconsistency when using the
older Karlsruhe value~\cite{kh80} 14.3$\pm$0.3.

For the s-wave scattering lengths we obtain $3a_{\pi^-p}=0.261$
m$_{\pi}^{-1}$ and $3a^{-}_{0+}=0.260$ m$_{\pi}^{-1}$, with 1-2\%
uncertainties. The $\pi^{-}p$ scattering length agrees with the PSI pionic
hydrogen result $3a^{\mbox{psi}}_{\pi^-p}=0.2649\pm0.0024$ m$_{\pi}^{-1}$,
while the isovector scattering length satisfies the
Goldberger-Miyazawa-Oehme (GMO) sum rule\cite{gol58} when using our
coupling constant and integral $J_{\mbox{gmo}}=-1.08\pm0.03$ mb$^{-1}$. The
p-wave scattering volume $a^{+}_{1+}=0.133$ m$_{\pi}^{-3}$ is consistent
with recent analyses of low energy data~\cite{lePWA}, as expected since the
resonanct P$_{33}$ partial wave dominates the low energy data and
$a^{+}_{1+}$.

The dispersion relations are very well satisfied up to about 1 GeV, in
general much better than KH80. From the forward $C^{+}(\omega)$ and
$E^{+}(\omega)$ dispersion relations, we obtained the coefficients
$\bar{d}_{00}=-1.30$ m$_{\pi}^{-1}$ and $\bar{d}_{01}= 1.19$
m$_{\pi}^{-3}$, in perfect agreement with the results from the slope and
intercept of the $C^{+}(0,t)$ subtraction constants at $t=0$, shown in
Fig.~\ref{fig:c+Extrap}.  The equivalent $\bar{d}_{01}$ result from the
$E^{+}$ dispersion relation, with its $\sim l^3$ sensitivity to partial
waves, and the $C^{+}(\nu,t)$ dispersion relation, with its $\sim l$
sensitivity, supports the reliability of our higher partial waves.

Figure~\ref{fig:c+Extrap} shows a polynomial fit to $C^{+}(0,t)$
near $t=0$, from which $\bar{d}_{02}$, $\bar{d}_{03}$, and $\bar{d}_{04}$
were estimated. The $\bar{d}_{02}$ coefficient is in perfect agreement with
the Karlsruhe result~\cite{bible}, while the sum of the higher order terms
yield a curvature correction $\Delta_{D}>11$ MeV, in agreement with the
$\pi\pi$ dispersion relation result\protect\cite{glls88} $12\pm1$ MeV.
Moreover, the curve extrapolates to about -4 MeV at the ``Adler point''
($t$=m$_{\pi}^2$), consistent with expected corrections to the Adler
Consistency Condition~\cite{adl65}, where it would be identically 0 in the
chiral limit. Compatibility with t-channel dispersion relations and chiral
constraints gives us confidence in the reliability of our subthreshold
coefficient results.

From the above subthreshold coefficients, and the curvature correction from
Ref.~\cite{glls88}, our result for the sigma term is $\Sigma$=79$\pm$7
MeV, compared to the Koch value 64$\pm$8 MeV~\cite{koch}. Though
surprising, the result is readily explained by the new experimental
information. Table~\ref{tab:sigma_terms} shows the breakdown of $\Sigma_d$
into its dispersion relation terms for both the KH80 solution and our own.
With respect to KH80 result, the new PSI pionic hydrogen and deuterium
scattering length~\cite{schr01} $a^{+}_{0+}\sim0.000$ m$_{\pi}^{-1}$, which
we reproduce, causes a 7 MeV increase.  With a $\pi NN$ coupling constant
$g^2/4\pi \sim 13.7$~\cite{pav99b,deSw97}, $\Sigma_d$ increases by 6
MeV~\footnote{The above increases were also noted in
  Ref.\protect\cite{schr01}}. It is well known that the KH80 solution
overshoots the data on the left wing of the $\Delta$ resonance. Our
solution fits the available data much better than KH80, resulting in a
narrower $\Delta$ width. This leads directly to the 3 MeV increases in each
of the dispersion integrals shown in Table~\ref{tab:sigma_terms}.
Consequently there is sound experimental evidence to support our new
$\Sigma$ result.

%-----------------------------------------------------------------------------
\subsection{Systematic Checks}

\begin{table}[t]
  \begin{center}
    \begin{tabular}{|l|r||c|c|c||c|c|c|}
      \hline
      Solution & $\Sigma_{d}$~[MeV] = & ``$a^{+}_{0+}$~const. & Born
      & $\int$D$^{+}$ & ``$a^{+}_{1+}$''~const. & Born & $\int$E$^{+}$ \\
      \hline\hline
      KH80 & 50 = & -7 & +9 & -91 & +352 & -142 & -72 \\
      \hline
      FA01 & 67 = &  0 & +9 & -88 & +351 & -136 & -69 \\
      \hline\hline
      {\bf difference} & 17 = & ${\bf +7}$ & {\bf 0} & {\bf +3}
      & ${\bf -1}$ & ${\bf+6}$ & {\bf +3} \\
      \hline
    \end{tabular}
    \caption{\small Comparison of $\Sigma_d$ from the
      Karlsruhe solution KH80~\cite{kh80} and our recent solution FA01.
      The change in the $C^+$ subtraction constant ($a^{+}_{0+}$) term, the
      $E^+$ Born term, and both integral terms are consistent with
      expectations from, respectively, pionic atom data
      \protect~\cite{schr01}, the coupling constant $g^2/4\pi \sim 13.7$
      (see Ref.~\cite{deSw97}), and a narrower $\Delta$ resonance
      width.  Values are rounded. See text for details.  \normalsize}
    \label{tab:sigma_terms}
  \end{center}
\end{table}

%%%%%%%%% Database Changes
Perhaps the most important systematic check is the sensitivity of our
results to the scattering database.  Around the $\Delta$ resonance, there
is a well known disagreement between the TRIUMF $\pi^{\pm}$ differential
cross section~\cite{triumf} and PSI $\pi^{\pm}$ total cross section
data~\cite{ped78} on the one hand, and the older CERN results~\cite{cern}
on the other. Only a small increase 0.07 and 4 MeV was observed in
$g^{2}/4\pi$ and $\Sigma_d$, respectively, for the ``CERN-only'' database.
In practice, both sets are included in the final fit\footnote{The other low
  and $\Delta$ resonance energy data are fit somewhat better in
  ``TRIUMF+PSI-only'' solution}. We found that weeding out large $\chi^2$
data sets had little effect on the result. Also, since the low energy data
are consistent with the PSI pionic atom results~\cite{schr01}, we conclude
that there are no large systematic effects from reasonable changes to the
current scattering database.

%%%%%%%%% Coulomb Corrections
The hadronic amplitudes are corrected for Coulombic effects following the
Nordita prescription~\cite{trom77}, supplemented in this analysis at high
energies by extended-source Coulomb barrier factors~\cite{gibb01}. The
current approach improved the agreement with the PSI pionic atom results
over our previous Nordita+point-source barrier results; however, neither
the coupling constant nor $\Sigma_d$ varied outside the errors when using
point- or extended-source barrier factors exclusively, or the Nordita
corrections supplemented by either. Moreover, the isospin-violating
$\Delta$ resonance is ``split'' defining ``hadronic''=``$\Delta^{++}$'',
consistent with the Nordita definition, but find no difference to our
previous approach with ``hadronic''=``$(\Delta^{0}+\Delta^{++})/2$'', or
with no splitting at all. We conclude that there are also no large
systematic uncertainties from our Coulomb correction scheme.

%%%%%%%%% Dispersion Relation Constraints
The implementation of our dispersion relation constraints was also checked.
We found that every reasonable form for the high energy amplitudes ($>$2
GeV) yields virtually identical results.  Agreement between the forward
subtracted and fixed-t unsubtracted dispersion relations is good for
reasonable constraints (\textit{i.e.} typical experimental error
$\sim$2\%), but suffers if they become too tight, $<$0.5\%. Constraining
the low energy $P_{13}$ partial wave to follow the Chew-Low form lowered
$\Sigma_d$ by 6 MeV, but once corrected, reasonable deviations caused changes
much smaller than our error bar. We also had solutions where the low
energy F and higher partial waves were not rigorously constrained to the
Koch values~\cite{koch86}, and no significant difference was found.
Furthermore, Olsson~\cite{olss00}, from a new dispersion relation sum rule,
and Kaufmann and Hite~\cite{kauf99}, from an interior dispersion relation
analysis, obtained values for $\Sigma$ consistent with our own using an
earlier SAID solution. Consequently, we are confident in the reliability of
our dispersion relation analysis.

%-----------------------------------------------------------------------------
\section{Summary}

In summary, we have performed a comprehensive partial wave and dispersion
relation analysis of the available $\pi N$ scattering data up to 2.1 GeV
that includes several improvements upon prior
analyses~\cite{vpiPapers,pav99}. For the pion nucleon coupling constant we
obtained $g{^2}/4\pi = 13.69\pm0.07$, consistent with our previous
determinations~\cite{vpiPapers,pav99b} and the Nijmegen
results~\cite{deSw97}. Our s-wave scattering lengths agree with the latest
PSI pionic hydrogen and deuterium results~\cite{schr01}. Our $\pi N$ sigma
term result is $\Sigma_d = 67\pm6$ MeV, or $\Sigma = 79\pm7$ MeV, compared
to the canonical result $64\pm8$MeV from Koch~\cite{koch}. These results
have proven robust with respect to the many systematic checks that we have
performed. In light of the large nucleon strangeness content $y/2\sim0.23$
inferred in the standard picture, we believe that alternate interpretations
of a large sigma term ought to be examined carefully.

%%%%%%%%%%%%%%%%%%%%%%%%%%%%%%%%%%%%%%%%%%%%%%%%%%%%%%%%%%%%%%%%%%%%%%%%%%%%%%%%%%
\acknowledgments{We gratefully acknowledge a contract from Jefferson Lab
  under which this work was done. The Thomas Jefferson National Accelerator
  Facility (Jefferson Lab) is operated by the Southeastern Universities
  Research Association (SURA) under DOE contract DE-AC05-84ER40150. We are
  grateful to W. Gibbs for the calculation of the extended source Coulomb
  barrier factors. M.M.P. thanks D. Bugg for the wine, and he and M.
  Sainio, G. Oades, W. Gibbs and B. Loiseau for fruitful discussions.  }

%%%%%%%%%%%%%%%%%%%%%%%%%%%%%%%%%%%%%%%%%%%%%%%%%%%%%%%%%%%%%%%%%%%%%%%%%%%%%%%%%%
%____________________________________________________________
%  Start references here:

\end{document}